\documentclass[conference]{IEEEtran}
\IEEEoverridecommandlockouts
\usepackage{cite}
\usepackage{amsmath,amssymb,amsfonts}
\usepackage{algorithmic}
\usepackage{graphicx}
\usepackage{textcomp}
\usepackage{xcolor}
\usepackage{booktabs}
\usepackage{tabularx}
\usepackage{subcaption}
\usepackage{multirow}
\usepackage{hyperref}
\usepackage{tikz}
\usepackage{microtype}
\usepackage{pifont}

\usepackage{tabularx,array}
\newcolumntype{C}{>{\centering\arraybackslash}X} 

\usepackage{array,adjustbox} 
\newcolumntype{M}[1]{>{\centering\arraybackslash}m{#1}} 
\usepackage{threeparttable}

\def\BibTeX{{\rm B\kern-.05em{\sc i\kern-.025em b}\kern-.08em
    T\kern-.1667em\lower.7ex\hbox{E}\kern-.125emX}}

\begin{document}
\bstctlcite{CTL:BSTcontrol}

\title{CAMformer: Associative Memory is All You Need}



\author{
\IEEEauthorblockN{
Tergel Molom-Ochir\textsuperscript{$\dagger$}*, 
Benjamin F. Morris\textsuperscript{$\dagger$}, 
Mark Horton\textsuperscript{$\dagger$}, 
Chiyue Wei\textsuperscript{$\dagger$}, 
Cong Guo\textsuperscript{$\dagger$}, 
Brady Taylor\textsuperscript{$\dagger$},\\
Peter Liu\textsuperscript{$\dagger$}, 
Shan X. Wang\textsuperscript{$\ddagger$}, 
Deliang Fan\textsuperscript{$\S$}, 
Hai ``Helen'' Li\textsuperscript{$\dagger$}, 
and Yiran Chen\textsuperscript{$\dagger$}
}
\IEEEauthorblockA{\textsuperscript{$\dagger$}Duke University, \textsuperscript{$\ddagger$}Stanford University, \textsuperscript{$\S$}Arizona State University}
\IEEEauthorblockA{\{tergel.molom-ochir, benjamin.morris, hai.li, yiran.chen\}@duke.edu}

\thanks{*Corresponding Author. This work was supported by the U.S. Department of Energy under Award Nos. DE-SC0026254 and SC0026382, 
NSF under Award Nos. 2328805 and 2328712, 
and AFOSR under Award No. FA9550-24-1-0322.}
}


\maketitle
\begin{abstract}
Transformers face scalability challenges due to the quadratic cost of attention, which involves dense similarity computations between queries and keys. 
We propose CAMformer, a novel accelerator that reinterprets attention as an associative memory operation and computes attention scores using a voltage-domain Binary Attention Content Addressable Memory (BA-CAM).
This enables constant-time similarity search through analog charge sharing, replacing digital arithmetic with physical similarity sensing. CAMformer integrates hierarchical two-stage top-$k$ filtering, pipelined execution, and high-precision contextualization to achieve both algorithmic accuracy and architectural efficiency.
Evaluated on BERT and Vision Transformer workloads, CAMformer achieves over 10$\times$ energy efficiency, up to 4$\times$ higher throughput, and 6–8$\times$ lower area compared to state-of-the-art accelerators—while maintaining near-lossless accuracy.
\end{abstract}


\section{Introduction}
Transformer-based models have become foundational in various domains, including natural language processing, computer vision, and speech recognition \cite{b1, transformer_survey}. 
Their ability to model long-range dependencies through self-attention mechanisms has led to significant performance improvements across numerous tasks \cite{transformer_genomics, vision_transformer, transformer_complex_systems, transformer_machine_translation, transformer_retrosynthesis, transformer_NLP, transformer_biomedicine}. 
However, the computational and memory demands of the attention mechanism, particularly its quadratic complexity with respect to sequence length, pose challenges for deploying Transformers in resource-constrained environments and for processing long sequences efficiently \cite{transformer_complexity}. 
Traditional hardware accelerators often address this bottleneck by optimizing matrix multiplication (MatMul) operations, such as the QK$^\top$ and AV computations inherent in the attention mechanism \cite{transformer_HW_survey, transformer_HW_survey_in_speech}.
Techniques like low-precision arithmetic, sparsity exploitation, and memory tiling have been employed to mitigate these challenges \cite{QUIDAM, EIE, SOFA}. 
Despite these efforts, the fundamental issue remains: the attention mechanism's reliance on dense matrix operations leads to substantial data movement and energy consumption. 
An alternative perspective considers attention as a form of content-based memory retrieval, akin to operations performed by content-addressable memory (CAM) systems \cite{DynCAM}.


Content-Addressable Memory (CAM), also known as associative memory, is a specialized form of memory that enables data retrieval based on content rather than specific addresses \cite{cam_survey_1, cam_survey_2}. Unlike traditional Random Access Memory (RAM), where data is accessed by providing an explicit address, CAM allows for simultaneous comparison of input data against all stored entries, returning the address of matching data \cite{FeFET_CAM, my_survey, CAM_AM}. This parallel search capability makes CAM exceptionally efficient for applications requiring rapid data matching and retrieval, such as IP look up, forwarding engine, Deep Random Forest acceleration, and DNA classification \cite{app_iplookup, app_forwarding_engine, app_DNA, app_DRF}. CAM operates by broadcasting the input data across the memory array, where each cell compares the input with its stored content. If a match is found, the corresponding address is returned. This mechanism eliminates the need for sequential searches, significantly reducing search time and enhancing performance in data-intensive tasks. By aligning the computational model of attention with the associative retrieval capabilities of CAMs, we can explore hardware architectures that inherently support efficient attention computations.

\begin{figure}[t]
    \centering
    \includegraphics[width=0.9\linewidth]{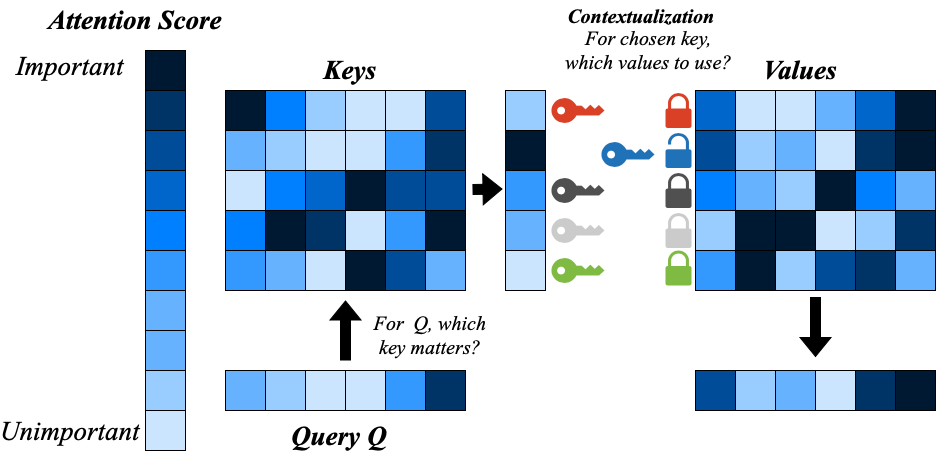}
    \caption{Attention as a key-lock mechanism. The query vector (Q) is used to determine similarity with stored keys (K), producing an attention score. The resulting soft selection determines which value vectors (V) to aggregate. This metaphor illustrates attention as an associative memory operation, where queries “unlock” relevant stored information.}
    \label{fig:key_lock_attention}
\end{figure}

\begin{figure*}[t]
    \centering
    \includegraphics[width=0.83\linewidth]{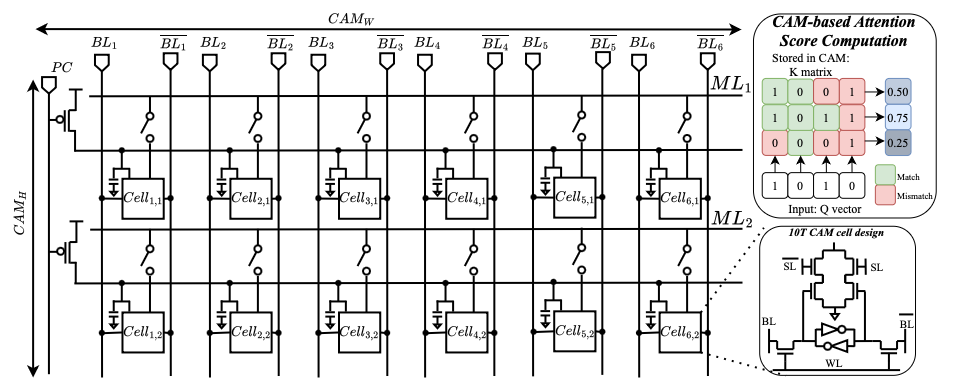}
    \caption{Array-level architecture of an example 2×6 BA-CAM module's array used for binary attention computation. Each row in the array performs parallel similarity matching against the broadcast query, with charge sharing accumulating match strength on shared matchlines. The inset shows the 10T1C CAM cell structure and an illustrative example of binary attention scoring based on Hamming similarity.}
    \label{fig:camformer-circuit}
\end{figure*}

\begin{figure}[b]
    \begin{minipage}{0.5\textwidth}%
      \begin{subfigure}[t]{0.55\linewidth}
        \centering
        \includegraphics[width=\linewidth]{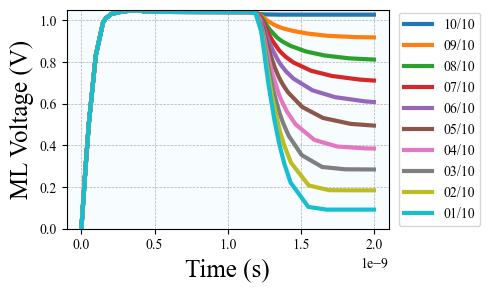}
        \caption{}
        \label{subfig:circuit1}
      \end{subfigure}
      \begin{subfigure}[t]{0.44\linewidth}
        \centering
        \includegraphics[width=\linewidth]{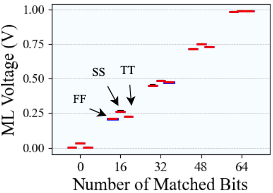}
        \caption{} 
        \label{subfig:circuit2}
      \end{subfigure}
    \end{minipage}%
    \caption{(a) Matchline voltage traces for varying partial matches in 1×10 BA-CAM. (b) PVT analysis across corners for 16×64 array.}
\end{figure}

In this work, we introduce CAMformer, a novel hardware accelerator that leverages CAM structures to perform attention operations. CAMformer reinterprets the attention mechanism as an associative memory process, enabling in-memory computation of attention scores. This approach improved throughput, lowers energy consumption, and offers scalability advantages. Our contributions are as follows:
\begin{itemize}
    \item \textbf{Reconceptualization of Attention}: We present a novel interpretation of the attention mechanism as an associative memory operation, aligning it with CAM functionalities.
    \item \textbf{CAM-based Attention Score}: We introduce a attention-score module that lowers circuit complexity for similarity search.
    \item \textbf{CAMformer Architecture}: We design a hardware accelerator that integrates CAM structures to perform attention computations, reducing reliance on traditional MatMul.
\end{itemize}

The remainder of this paper is organized as follows: 
Seciton~\ref{sec:binary-attention-cam} details the CAMformer circuits and the MatMul operation.
Section~\ref{sec:arch} details the CAMformer architecture and its operational principles. Section~\ref{sec:results} presents performance evaluations and comparisons. Finally, Section~\ref{sec:conclusion} concludes the paper and outlines potential directions for future research.


\section{Binary Attention CAM}
\label{sec:binary-attention-cam}

\subsection{Circuit-Level Design}
\subsubsection{Cell design}
We implement a 10T1C (10 transistors 1 capacitor) CAM cell tailored for partial-match, and binary vector-matrix multiplication operations. 
Each cell stores 1-bit data in SRAM logic, and its match results representation in a $22fF$ MIM capacitor, and compares it to the query via XNOR logic.
When matched, the precharged capacitor stays high; otherwise, discharged. 
The charge-sharing mechanism along the matchline enables analog accumulation of Hamming similarity, replacing digital popcount logic \cite{CapCAM, ASMCap}. 
The design operates in four phases—precharge, broadcast, match, and charge share—and avoids destructive readout while supporting pipelined operation.


\subsubsection{Array \& Matchline Architecture}
The BA-CAM array computes binary VMM by broadcasting the query of $Q\in\{0,1\}^{d_k\times 1}$ across columns of $K^\top\in\{0,1\}^{d_k\times N}$ and accumulating bitwise match results as analog voltages in $[0,1]$ on each matchline, resulting in $A\in[0,1]^{1\times N}$. 
These voltages, linearly proportional to Hamming similarity, are digitized using shared 6-bit SAR ADCs. 
Unlike TD-CAM, which uses time-domain delay sensing and requires time-difference amplifiers (TDA), our voltage-based scheme is simpler, faster, and significantly more robust to PVT variations \cite{td_cam}. 
This linear, delay-free sensing model eliminates timing calibration and scales efficiently to larger arrays and higher throughput without analog complexity.
BA-CAM's array is shown in Fig. \ref{fig:camformer-circuit}.
Compared to the conventional exact matching operation, partial matching via charge sharing allows Hamming distance computation in analog (shown in Fig. \ref{subfig:circuit1}).

\subsection{Microarchitecture \& VMM Engine  }
\subsubsection{BIMV — Binary In-Memory Vector-Matrix Multiplication}
\label{subsubsec:bimv}

We implement binary $QK^\top$ with a \emph{Binary Matrix–Vector (BIMV)} engine built on a \emph{Binary-Attention CAM (BA-CAM)} array. Each row stores a binary key (from $K^\top$); the query is broadcast. Bitwise \texttt{XNOR} happens in parallel and matching bits charge-share onto the matchline; the resulting voltage encodes Hamming similarity and is sensed by a simple ADC/comparator, eliminating digital arithmetic and yielding constant-latency compute~\cite{CapCAM}. CAM performs similarity on the matchline (time/voltage), whereas CiM performs \texttt{XNOR}+popcount on bit-lines and digitizes via column-muxed ADCs—adding peripheral/serialization overhead (Table~\ref{tab:cam_bimv_comparison}). We extend capacitive CAM for binary MatMul by linearly scaling matchline outputs with a 6-bit ADC: $s=2\cdot\mathrm{ADC}(v)-CAM_W$, mapping $[0,1]\rightarrow[-64,64]$ while preserving attention-score ordering. Unlike digital BIMV (e.g., XNOR-NE~\cite{xnor_ne}, BitFusion~\cite{bitfusion}) that needs SRAM fetch, external \texttt{XNOR}–popcount, and sequential accumulation, CAM unifies compute and memory. TD-CAM~\cite{td_cam} removes digital popcount and is row-parallel, but encodes scores in discharge \emph{delay}, requiring TDAs, tight delay matching, and calibration—introducing timing complexity and weakening robustness.

\begin{table}[t]
\small
\caption{Circuit-Level Comparison of BIMV / Attention-Score \emph{Modules}}
\label{tab:cam_bimv_comparison}
\centering
\begin{adjustbox}{max width=\columnwidth} 
\begin{tabular}{|M{1.6cm}|M{2.5cm}|M{2.2cm}|M{2.35cm}|}
\hline
\textbf{Feature} & \textbf{CiM [29]} & \textbf{TD-CAM [28]} & \textbf{BA-CAM (Ours)} \\\hline
Sensing      & BL sum (XNOR+Accumulate)        & Time ML          & Voltage ML \\\hline
Similarity   & No (popcount)         & Yes (delay)      & Yes (voltage) \\\hline
Peripherals  & Flash ADC (MUX) + Adder Tree & TDA + tune       & Shared SAR \\\hline
Tech         & 65\,nm              & 65\,nm           & 65\,nm \\\hline
Module area  & High (ADC)          & Med--High (TDA)  & \textbf{Low} (shared SAR) \\\hline
$V_{\!DD}$   & 0.6--1.0\,V         & 1.2\,V           & 1.2\,V \\\hline
Freq         & 18.5\,MHz           & 200\,MHz         & \textbf{500\,MHz} \\\hline
Overall err. & 7\% (pred.)         & 7.76\%           & \textbf{1.12}\%$^{\star}$ \\\hline
PVT robustness         & Moderate (Calibrated ADC)    & Low (time-domain)     & \textbf{High} (voltage-sensed) \\\hline
Complexity   & Very high (ADC+Adder Tree)     & High (TDA)       & \textbf{Low} (no MAC/popcnt) \\\hline
\end{tabular}
\end{adjustbox}

\footnotesize $^{\star}$ simulated at $\sigma=1.4\%$.
\end{table}

\begin{figure}[b]
    \centering
    \includegraphics[width=0.9\linewidth]{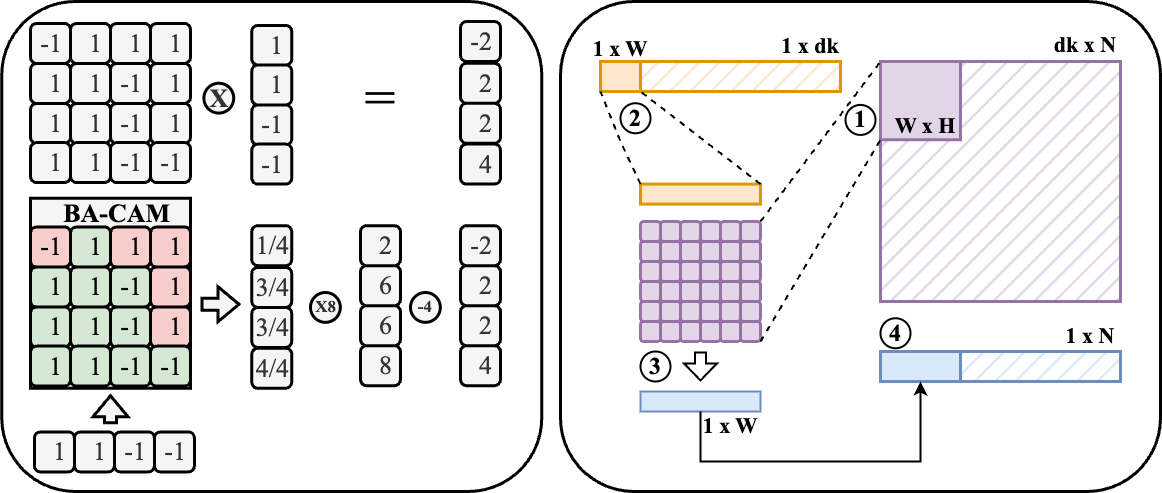}
    \caption{Illustration of matrix-vector multiplication. Comparison of conventional (left top) versus CAM-based (left bottom). Tiling steps for larger matrix-vector operations (right).}
    \label{fig:multiplication}
\end{figure}


Our BA-CAM encodes similarity by \emph{voltage}, not time: each matching bit adds charge to the matchline; the sensed voltage directly reflects similarity, removing delay-path tuning and TDAs. Unlike TD-CAM’s nonlinear delay, BA-CAM yields a linear voltage response, stronger PVT tolerance, and cleaner digital integration (Table~\ref{tab:cam_bimv_comparison}). It scales flexibly and needs no majority-threshold voting. Under PVT, BA-CAM holds matchline deviation within $5.05\%$ with mean error as low as $1.12\%$ across TT/SS/FF (Fig.~\ref{subfig:circuit2}), outperforming prior TD-CAMs that show delay deviations up to $\le 7.76\%$. BA-CAM avoids timing-based sensing entirely—computing similarity physically (via voltage) in constant time—eliminating external logic, alignment, and calibration, and delivering energy-efficient, fully in-memory associative compute; digital BIMV \emph{emulates} association, BA-CAM \emph{is} associative.

We illustrate the operation of BIMV in BA-CAM in Fig.~\ref{fig:multiplication}.
On the top left, a conventional approach to BIMV is shown. 
Digital components perform multiplication using digital logic.
On the bottom left, a BA-CAM array performs the binary multiplication/matching and accumulation all in the analog domain, producing partial results in the range $[0,1]$.
These values are converted to signed values centered around $0$ by the fixed functional multiply and subtract units.

On the right, we illustrate how we use tiling to general BIMV for tensors with dimensions larger than the CAM array: $Q\in\{-1,+1\}^{1\times d_k}$ and $K^\top\in\{-1,+1\}^{d_k\times N}$.
We assume that $d_k$ and $N$ are multiples of $CAM_W$ and $CAM_H$ for simplicity, respectively, but padding can be applied otherwise.
In step \ding{192}, we load a $CAM_W\times CAM_H$ sized tile of $K^\top$ into the BA-CAM array.
In step \ding{193}, a $CAM_W$ sized segment of the $Q$ vector is loaded into the query register.
Then, in step \ding{194}, we perform the associative tiled-MAC operation to receive our partial output of size $CAM_W$.
If $N>CAM_W$, we tile horizontally and concatenate partial results to the final result vector in step \ding{195}.
If $K>CAM_H$, we must tile vertically and accumulate partial results into the same segment of the final result vector (repeat steps \ding{192}-\ding{195} again).

\begin{figure}[t]
    \centering
    \includegraphics[width=0.8\linewidth]{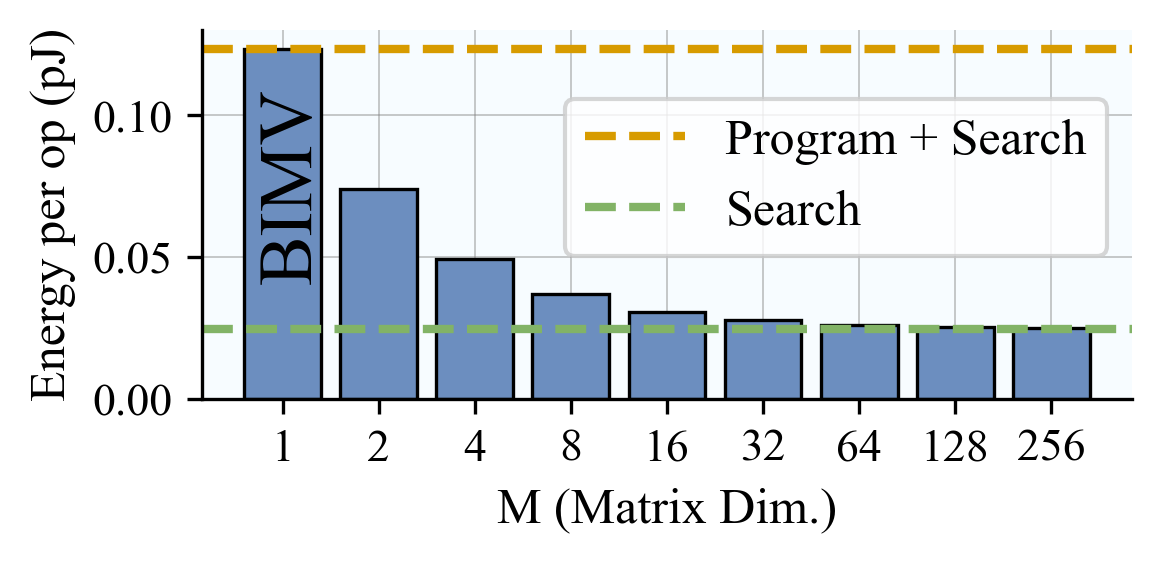}
    \caption{Per-op energy vs. matrix dimension $M$ in BA-CAM. Larger $M$ reduces energy by amortizing programming cost. Dashed lines show search-only and total energy bounds.}
    \label{fig:bimm}
\end{figure}

For higher-precision $V$, we decompose $K^\top$ entries into binary slices (LSB$\rightarrow$MSB) and run per-slice BIMM. Slice outputs are digitally shifted and accumulated, adding precision without changing the CAM path. This supports binary–integer MatMul and quantized $V\in{\texttt{int2},\texttt{int4},\texttt{int8}}$.

\begin{figure*}[tb]
    \centering
    \includegraphics[width=\linewidth]{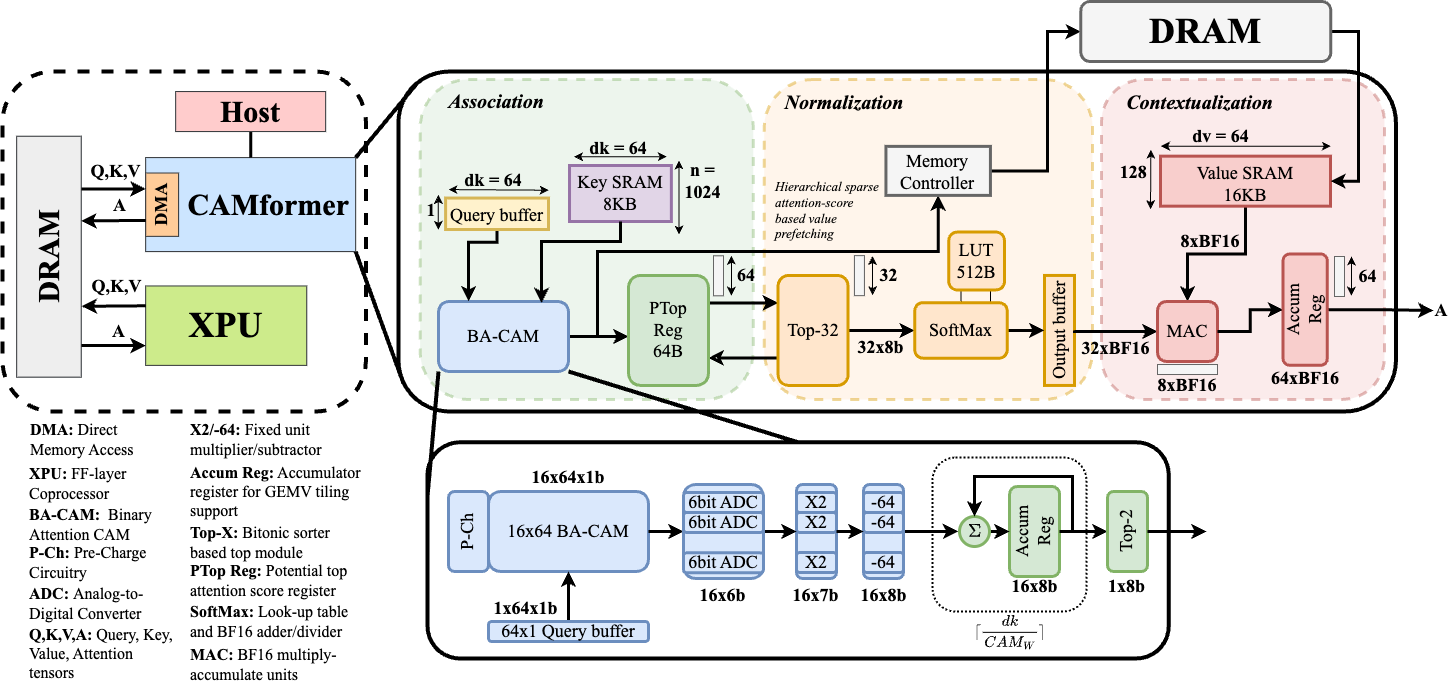}
    \caption{The design consists of three pipelined stages---association, normalization, and contextualization---centered around a 16×64 BA-CAM array. The BA-CAM computes binary attention scores, which are sparsified and normalized before BF16 attention output is computed using BF16 MACs. Integration with XPUs and DRAM enables efficient end-to-end attention processing.}
    \label{fig:camformer-arch}
\end{figure*}
\section{CAMformer Architecture}
\label{sec:arch}


In this section, we develop CAMformer, an attention accelerator built around the BA-CAM circuit introduced in Sec.~\ref{sec:binary-attention-cam}.
CAMformer attention is shown in Eq. \ref{eq:attention}.
First, we explain the execution model of CAMformer and explain how a larger Deep Learning (DL) system can integrate CAMformer into the attention workflow in Sec.~\ref{subsec:arch-systemintegration}.
With this model in mind, we next introduce the high-level CAMFormer architecture in Sec.~\ref{subsec:arch-camformeroverview}.
Finally, we explore in detail some of the novel optimizations that enable CAMformer to surpass state-of-the-art attention computation performance.

\begin{equation}
    \label{eq:attention}
    \begin{split}
        \text{CAMformer-Attn}(Q, K, V) \\= \text{SoftMax}&\big(\text{Top-32}( QK^T)\big) \cdot V
    \end{split}
\end{equation}

\subsection{System Integration}
\label{subsec:arch-systemintegration}
CAMformer operates as an attention accelerator for large-scale DL systems.
The CAMformer accelerator integrates into existing systems that contain other accelerated processing units, XPUs (GPUs, TPUs, etc.), responsible for the dense matrix multiplications such as FF (feed-forward) layers.
We illustrate the integration of CAMformer into a larger system in Fig.~\ref{fig:camformer-arch}.
We utilize shared memory-based communication with XPUs for the binary $Q,\,K$ tensors and BF16 $V,\,A$ tensors.
While an external host (CPU) programs and monitors the CAMformer accelerator, CAMformer uses a local Direct Memory Access (DMA) engine and memory controller for fast access to global memory during computation.
CAMformer is currently optimized for decoder-style (causal) attention.

\subsection{CAMformer Overview}
\label{subsec:arch-camformeroverview}

CAMformer has three pipelined stages (Fig.~\ref{fig:camformer-arch}): \emph{Association} computes $QK^\top$ via BA-CAM and launches hierarchical sparse ranking; \emph{Normalization} finalizes ranking and applies softmax, $\hat{A}=\mathrm{SoftMax}(\mathrm{Top\text{-}32}(QK^\top))$; \emph{Contextualization} performs high-precision sparse MV with $V$, yielding $A=\hat{A}V$. Each stage is separable, feed-through, and consumes/produces the prior/next stage’s data.

\subsubsection{Association}
We compute attention scores from binary $Q,K$. The Key SRAM stores the full binarized $K$ and is off the critical path since many queries reuse one $K$. The Query buffer holds a single query (batch=1). BA-CAM executes the binarized MVM (Sec.~\ref{subsubsec:bimv}). Although batching can improve $QK^\top$ energy, it inflates downstream hardware, so we use batch=1. The CAM is $16\times64$: height 16 reduces ADC overhead; width 64 avoids vertical tiling for $d_k{=}64$ (Sec.~\ref{subsubsec:bimv}). For larger $d_k$, an accumulation register enables vertical tiling. For each tile, we program BA-CAM and run an associative tiled MAC with the query; ADC precision covers the full match range. A bitonic Top-2 picks the highest score per tile; we keep it for sparse attention and drop others. The bitonic sorter also makes sparsity easily configurable. Per tile, Top-2 scores go to a potential-top register, and indices go to the local memory controller to prefetch the corresponding $V$ entries.
We use $g{=}16$ tiles with Top-2 per tile $\Rightarrow$ Top-32 overall.
$k$ fixes the returned indices (V-buffer depth), so we co-design $k{=}32$ with V-SRAM capacity: large enough to shrink candidates, small enough to bound accuracy loss; larger $k$ offers diminishing returns (cf. recall bound).
If stage-1 scores satisfy $|\hat s_i - s_i|\le\varepsilon$, a margin $\Delta_k \equiv s_{(k)}-s_{(k+1)} > 2\varepsilon$ ensures $\mathrm{recall@}k=1$. For binary similarity (mean of $m$ Bernoulli matches, e.g., $m=d_k$), Hoeffding yields $\Pr[\text{drop any true top-}k] \le k(N-k)\exp(-2m\,\delta_{\min}^2)$, consistent with HAD’s binarized Q/K + top-$N$ sparsification.~\cite{hoeffding1963,HAD}

\subsubsection{Normalization} We select the top-32 attention scores from the 128 candidates produced in association. A bitonic-sorter Top-32 block provides runtime sparsity flexibility. To reduce area, we use a 64-input module and refine across batches as each 16-tile group yields 32 new top-2 candidates (after the initial 32 tiles). The SoftMax engine uses a 512 B LUT, one BF16 accumulator, and one BF16 divider. For each of the 32 8-bit scores, we compute $\exp(x)/\sqrt{d_k}$ via the LUT, accumulate the denominator on the fly, and normalize once it is complete. The outputs are valid probabilities: each in $[0,1]$ and summing to 1.


\subsubsection{Contextualization} In the contextualization stage, we compute the BF16 Attention tensor.
The use of higher-precision BF16 has been noted as a requirement for maintaining model accuracy~\cite{HAD}.
Our Value SRAM has been loaded with a sufficient number of valid entries to begin the multiply-accumulate (MAC) operations.

\begin{figure}[t]
    \centering
    \includegraphics[width=0.81\linewidth]{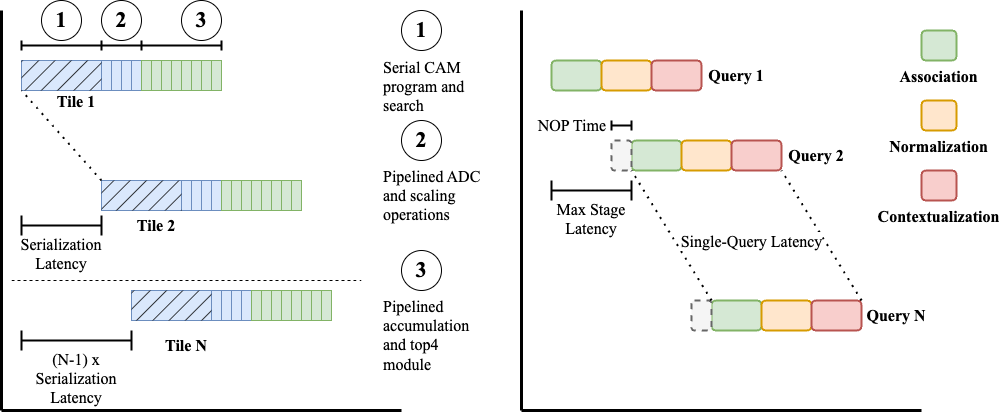}
    \caption{CAMformer pipelining strategies. Left: Fine-grained pipelining overlaps CAM operations within the association stage. Right: Coarse-grained pipelining enables query-level parallelism across all stages.}
    \label{fig:pipelining}
\end{figure}

\begin{figure}[b]
    \centering
    \includegraphics[width=0.86\linewidth]{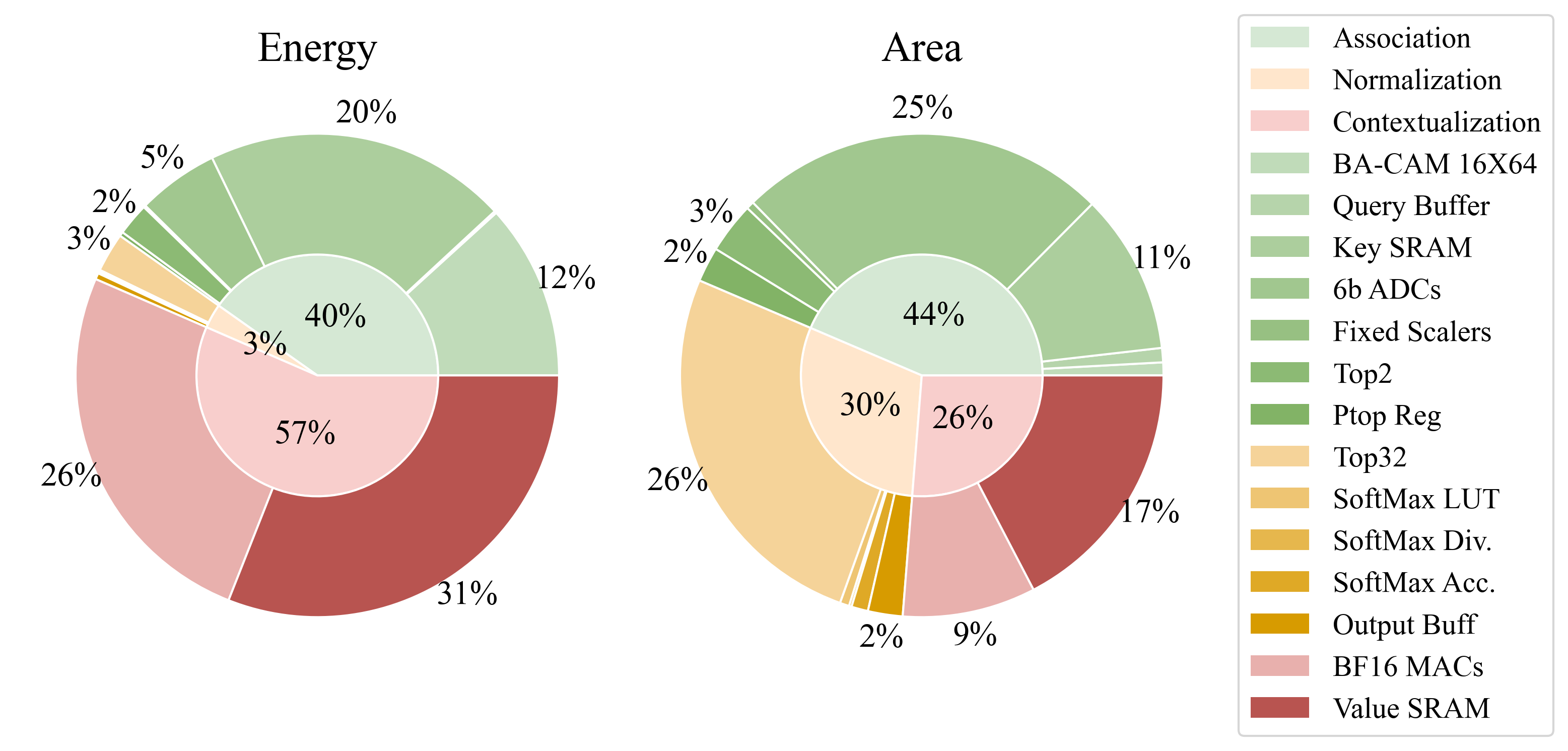}
    \caption{Breakdown of CAMformer energy and area. Energy is dominated by BF16 MACs and Value SRAM, while area is split across all stages with largest contributions from SRAM and normalization logic.}
    \label{fig:areapowerbreakdown}
\end{figure}

\subsection{Optimizations}
\label{subsec:arch-optimizations}

\subsubsection{Fully Binarized Attention-Score}
Fully binarizing $Q$ and $K$ enables analog $QK^\top$ via the associative BA-CAM (Sec.~\ref{sec:binary-attention-cam}) and cuts on-chip storage to 6.25\% of BF16 for the Query buffer and Key SRAM. The resulting bounded score range also makes SoftMax cheap: a small LUT handles $\exp(\cdot)$ and normalization.


\subsubsection{Fine-grained pipelining}

We utilize fine-grained pipelining to accelerate the critical path of each stage.
The association stage uses fine-grained pipelining strategies so that tiling steps in different parts of the pipeline can operate concurrently.
We illustrate the fine-grained pipelining and effect of CAM serialization latency requirements for the association stage in Fig.~\ref{fig:pipelining} left.
The normalization stage utilizes fine-grained pipelining in the SoftMax module.
Because the normalization stage is not on the critical path, we perform the accumulation and division serially.
By utilizing a pipelined BF16 divider, the overall latency for the SoftMax operation is greatly reduced from $32t_\text{div}$ to $31 + t_\text{div}$, where $t_\text{div}$ is the end-to-end latency of the BF16 divider.
In the contextualization stage, we utilize fine-grained pipelining for the MAC operations.

\subsubsection{Coarse-grained pipelining}
In addition to fine-grained pipelining within each stage, we also perform coarse-grained pipelining between stages.
This improves hardware utilization by ensuring that each stage remains busy.
The overall throughput is determined by the latency of the longest stage.
Other stages will incur an amount of stall time equivalent to the difference between their latency and the longest stage latency.
The coarse-grained pipelining strategy and corresponding total no-op time are illustrated in Fig.~\ref{fig:pipelining} right. 
In order to maximize hardware utilization, we balance the throughput of the stages in our design space exploration (Sec.~\ref{subsec:results-dse}).

\begin{figure}[t]
    \centering
    \includegraphics[width=0.85\linewidth]{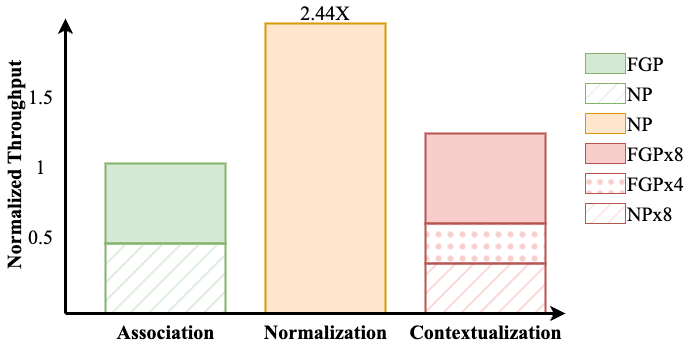}
    \caption{CAMformer throughput by stage. Fine-grained pipelining and parallelism boost association and contextualization stages, enabling balanced pipeline performance.}
    \label{fig:pipelining-dse}
\end{figure}

\subsubsection{Hierarchical sparse attention-score ranking}
We use a two-stage top-$k$ to cut on-chip score storage and enable $V$ prefetch. Stage-1 keeps top-2 per 16 during each tile; Stage-2 finalizes ranking. Each top-2 triggers the MC/DMA to fetch the corresponding $V$. $V$ is laid out contiguously in DRAM: rows of $64\times16\text{b}$, so $64$ rows fit an $8$KB page. With no interleaving, one $t_{\mathrm{RC}}$ serves each set of $64$ scores; using HBM3 $t_{\mathrm{RC}}=48$ns~\cite{dramsim3}, the pipeline fully hides DRAM latency. Required bandwidth is $\approx50$,GB/s, which a single HBM3 channel can sustain~\cite{hbm3jedec}.

\begin{table}[b]
\setlength{\tabcolsep}{3pt}
\caption{Performance Comparison of CAMformer Variants vs. Existing Architecture Solutions at 1\,GHz}
\label{tab:comparison_existing_arch}
\centering
\begin{threeparttable}
\begin{tabular}{l ccccccc}
\toprule
Accelerator & Q/K/V & Core & Thruput & Energy Eff. & Area & Power\\
            & bits  & (\#) & (qry/ms) & (qry/mJ) & (mm\textsuperscript{2}) & (W)\\
\midrule
MNNFast\cite{mnnfast}         & 32/32/32   & 1  & 28.4 & 284  & --   & 1.00$^{*}$\\
A\textsuperscript{3}\cite{A3} & 8/8/8      & 1  & 52.3 & 636  & 2.08 & 0.82\\
SpAtten\textsuperscript{1/8}\cite{spatten} & 12/12/12 & 1 & 85.2 & 904 & 1.55 & 0.94\\
HARDSEA\cite{hardsea} & 8/8/8 & 12 & 187$^\dagger$ & 191$^\dagger$ & 4.95 & 0.92 \\
CAMformer                     & 1/1/16     & 1  & 191  & 9045 & 0.26 & 0.17\\
CAMformer\textsubscript{MHA}  & 1/1/16     & 16 & 3058 & 9045 & 4.13 & 2.69\\
\bottomrule
\end{tabular}

\begin{tablenotes}[flushleft]
\footnotesize
\item \textbf{Notes:} MNNFast does not report Q/K/V precision; treated as FP32 (32/32/32). 
SpAtten uses progressive MSB{+}LSB quantization (e.g., 6{+}4), with an up-to-12b on-chip datapath. $^\dagger$Converted from 802.1 GOPS and 821.3 GOPS/W assuming 4.3 GOP/query.
\end{tablenotes}
\end{threeparttable}
\end{table}

\section{Results}
\label{sec:results}

\subsection{Experimental Methodology}
We implement CAMformer’s digital blocks in Verilog and synthesize with Synopsys Design Compiler (TSMC 65 nm) for area/timing/power. Analog CAM arrays are characterized in HSPICE. ADC, BF16 MAC, and BF16 divider costs follow \cite{6badcs,bf16mac,bf16div} and are scaled to 45 nm via \cite{scaling}; DRAM energy is 2.33 nJ/bit \cite{memorypower}. A Python system simulator models performance, energy, and area. Metrics include end-to-end attention latency (with memory), energy (GOP/J), power, and queries/s. CAMformer\textsubscript{MHA} spans 16 heads across all 16 HBM channels. We compare against TPU \cite{tpuv4}, WSE \cite{wse2}, SpAtten \cite{spatten}, and A$^3$ \cite{A3} (Table~\ref{tab:comparison_existing_arch}).

\subsection{Design Space Exploration}
\label{subsec:results-dse}

We balance the throughput of CAMformer's three stages through fine-grained pipelining and data parallelism optimizations (Fig.~\ref{fig:pipelining-dse}). The normalization stage provides sufficient throughput with minimal parallelism due to sparse attention optimization, while the contextualization stage requires 8 parallel MAC units to match the association stage's throughput. As shown in Fig.~\ref{fig:areapowerbreakdown}, area is distributed across SRAM (42\%), Top-32 module (26\%), and processing units, while energy is dominated by the contextualization stage (57\%) due to BF16 precision requirements; component-wise, Value/Key SRAM account for 31\%/20\%, MACs 26\%, and BA-CAM 12\% of total energy.

\begin{table}[b]
    \centering
    \caption{Top-1 accuracy with two-stage HAD on DeiT models. Accuracy remains near baseline for $k \geq 2$ with minimal degradation using group size 16.}
    \begin{tabular}{lrrr}
        \toprule
        first stage k & DeiT‑B & DeiT‑S & DeiT‑T \\
        \midrule
        HAD baseline & 79.24 & 75.60 & 66.58 \\
         k=8 & 79.27 & 75.68 & 66.53 \\
         k=4 & 79.26 & 75.65 & 66.48 \\
         k=2 & 79.16 & 75.29 &  65.86 \\
         k=1 & 78.11 & 72.32 & 61.03 \\
        \bottomrule
    \end{tabular}
    \label{tab:imagenet_res}
\end{table}

\subsection{State-Of-The-Art Comparison}
\begin{figure}[t]
    \centering
    \includegraphics[width=0.85\linewidth]{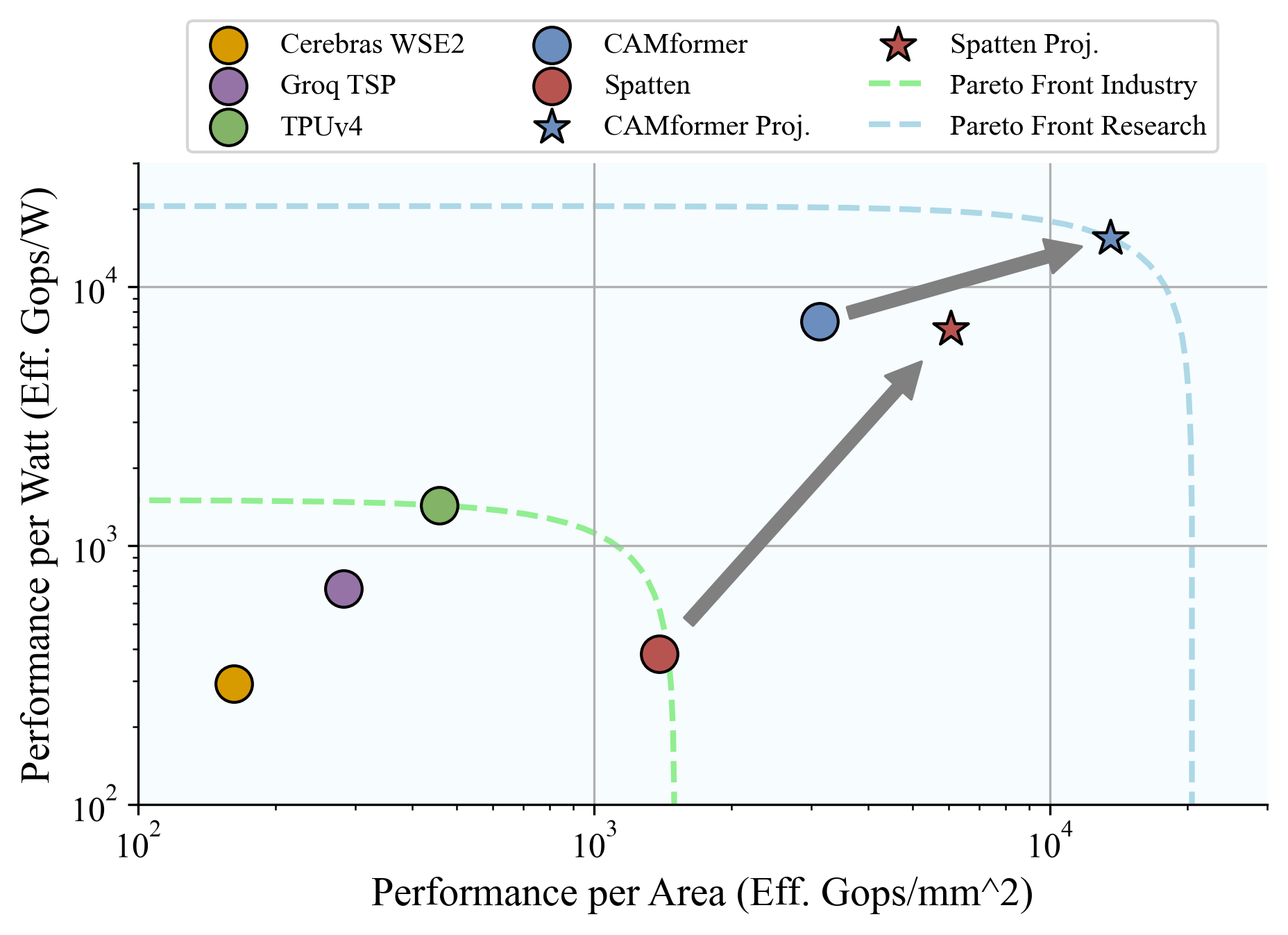}
    \caption{CAMformer (and projected scaling) lies on the research Pareto frontier, surpassing TPUv4 and WSE2 in performance-per-area and performance-per-watt; points report \emph{effective} GOPS/W at the Table~\ref{tab:comparison_existing_arch} Q/K/V precisions under fixed accuracy/latency, not peak TOPS.}
    \label{fig:pareto}
\end{figure}

We compare performance of state-of-the-art accelerators on single query attention processing for a BERT-Large~\cite{bert} model with 16 heads, $d_k = d_v = 64$, and sequence length $n=1024$.
The same CAM top-$k$ mechanism extends to decoders - CAM search over a growing KV cache each step (causal) - and to encoder–decoder models via non-causal search over encoder keys.
For longer contexts, CAMformer would scale by provisioning proportionally larger BA-CAM (keys) and V-SRAM (values) sized to the target maximum context (the per-step top-$k$ V-buffer remains fixed by $k$); note that KV-cache memory grows with sequence length in practice.
In Fig.~\ref{fig:pareto}, we compare industry products: Cerebras WSE2~\cite{wse2}, Groq TSP~\cite{tsp}, Google TPUv4~\cite{tpuv4} to the best-performing academic accelerators on this attention workload.
We find that the Pareto front of projected academic accelerators (projected node from 45nm to 22nm), defined at the CAMformer point, exceeds that of the industry products, defined at the TPUv4 point.

\subsection{Algorithmic Accuracy}

\begin{table}[b]
    \centering
    \caption{GLUE accuracy with two-stage HAD using group size 16. Accuracy remains comparable to the single-stage baseline for $k = 2$ and $k = 4$, with less than 0.4\% average degradation.}
    \begin{tabular}{lrrr}
    \toprule
    Metric       & HAD baseline   & first‑stage k=4  & first‑stage k=2  \\
    \midrule
    MNLI         & 82.45/82.84    & 82.37/82.98   &  82.31/82.74 \\
    QQP          & 90.11          & 90.01         &  89.87 \\
    QNLI         & 89.68          & 89.60         &  89.54 \\
    SST‑2        & 91.63          & 91.42         &  91.28 \\
    CoLA         & 55.47          & 55.16         &  54.90 \\
    STS‑B        & 87.46          & 87.27         &  87.27 \\
    MRPC         & 83.82          & 83.87         &  83.87 \\
    RTE          & 65.70          & 64.33         &  64.62 \\
    \hline
    Avg          & 80.81          & 80.54         &  80.48 \\
    \bottomrule
    \end{tabular}

    \label{tab:glue_res}
\end{table}


Hamming Attention Distillation (HAD)~\cite{HAD} binarizes $Q,K$ with $< 3\%$ top-1 drop on ImageNet~\cite{imagenet} and GLUE~\cite{glue}. CAMformer replaces single-stage top-$k$ with two-stage top-$k$ for early filtering and prefetching, reducing storage and hiding DRAM latency. Added loss is negligible: ImageNet $\le 0.2\%$ even with small first-stage $k$ (Table~\ref{tab:imagenet_res}); GLUE (group $=16$, $k\in{2,4}$) $\le 0.4\%$ vs. single-stage HAD (Table~\ref{tab:glue_res}). Thus, CAMformer keeps HAD’s accuracy while enabling efficient binary-attention inference.



\section{Conclusion}
\label{sec:conclusion}
We presented CAMformer, a novel accelerator that reinterprets attention as an associative memory operation. 
Using a voltage-domain Binary Attention CAM (BA-CAM), CAMformer computes similarity scores with constant latency and minimal overhead. 
Our design achieves over 10$\times$ energy efficiency, up to 4$\times$ higher throughput, and 6–8$\times$ lower area compared to state-of-the-art attention accelerators. while maintaining near-lossless accuracy via two-stage Hamming Attention Distillation.


\clearpage

\bibliographystyle{IEEEtran}
{\tiny 



}


\end{document}